\documentclass[journal=jacsat,manuscript=article]{achemso}

\usepackage[version=3]{mhchem} 

\author{Karel L. K. De Witte}
\author{Tom Braeckevelt}
\author{Massimo Bocus}
\author{Sander Vandenhaute}
\author{Veronique Van Speybroeck}
\affiliation{Center for Molecular Modeling, Ghent University, 9052 Zwijnaarde, Belgium}
\email{Veronique.VanSpeybroeck@UGent.be}

\title[NPT TI scheme]
  {A Novel NPT Thermodynamic Integration Scheme to Derive Rigorous Gibbs Free Energies for Crystalline Solids}

\SectionNumbersOn
\begin{document}
\begin{tocentry}

\centering
\includegraphics[width=1.0\linewidth]{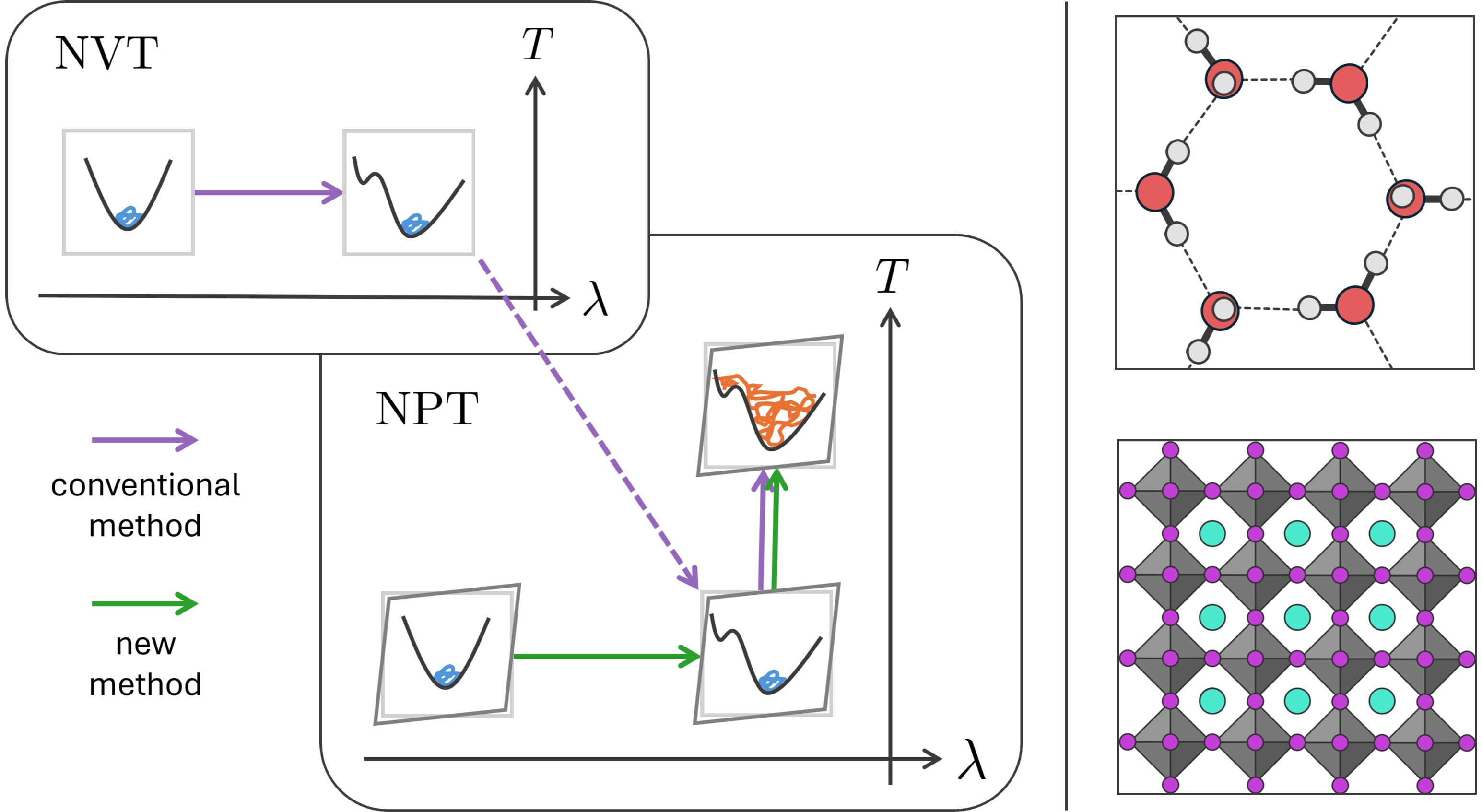}

\end{tocentry}

\begin{abstract}

Thermodynamic Integration (TI) is the state-of-the-art computational technique for accurate Gibbs free energy predictions of solids. Conventional TI schemes start from an NVT harmonic reference and require three successive corrections to recover the Gibbs free energy of the real crystal in the NPT ensemble. However, the NVT-to-NPT correction neglects full cell flexibility. Here, we present a rigorous (and only) two-step TI scheme that operates entirely in the NPT ensemble, eliminating the need for the approximate NVT-to-NPT step. The key methodological advancement is the novel NPT reference that explicitly accounts for full cell fluctuations. The new approach is compared with the conventional one via two complementary case studies. For ice polymorphs, having simple cell-shape distributions, the new approach reproduces conventional TI results with excellent agreement. For \ce{CsPbI3}, whose black phase exhibits complex cell-shape behavior, we demonstrate that our novel method provides more accurate Gibbs free energy differences than the conventional one. Moreover, the proposed framework maintains comparable computational cost while offering a simplified workflow. Overall, the new NPT TI scheme provides rigorous and direct Gibbs free energy calculations for solids.

\end{abstract}


\section{Introduction}

\begin{figure}
    \centering
    \includegraphics[width=1.0\linewidth]{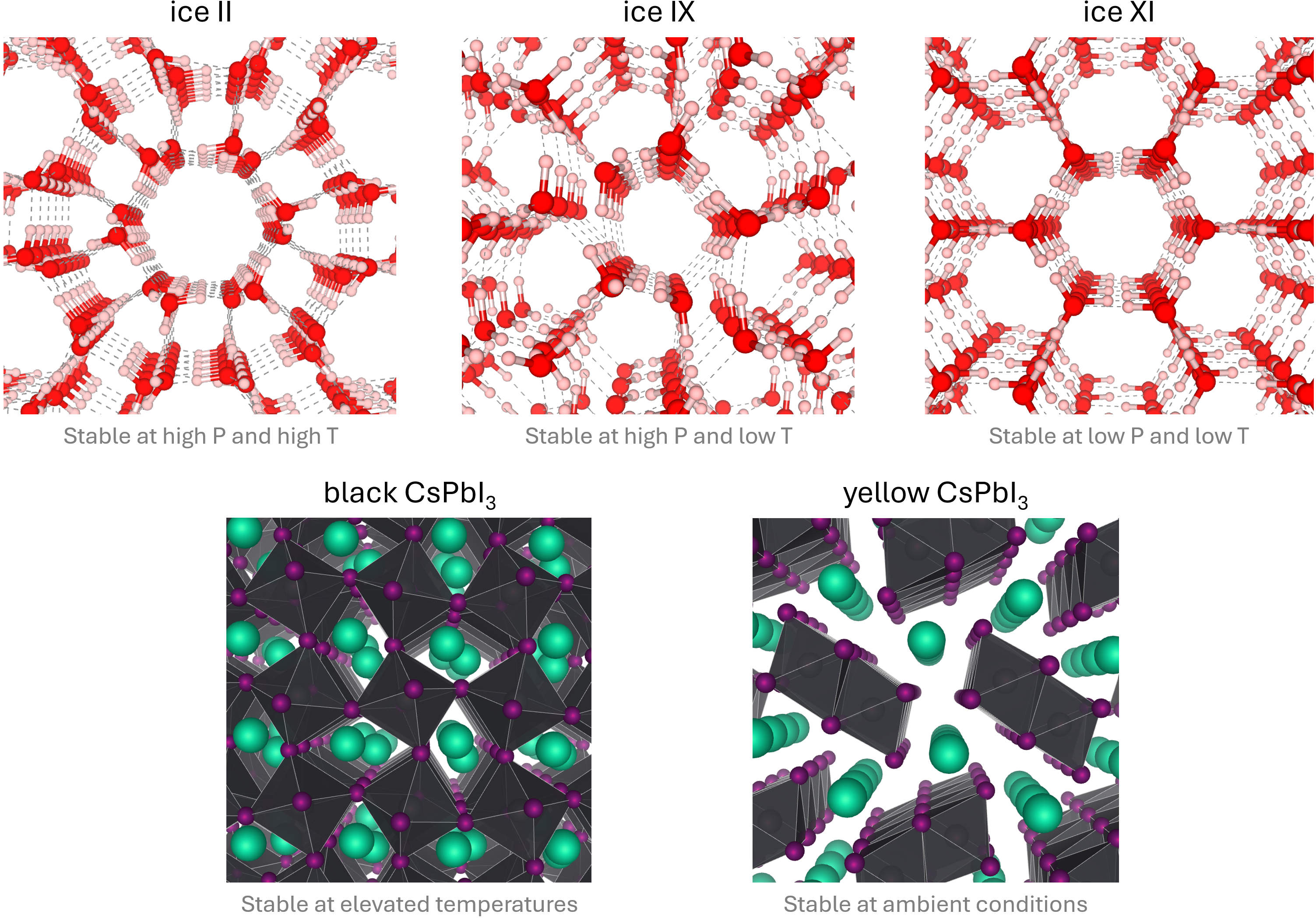}
    \caption{\label{fig:materials}\textbf{Overview of the case study materials.} The first case study covers three ice phases (red atoms indicate oxygen; white atoms indicate hydrogen). The second case study comprises the black and yellow phase of \ce{CsPbI3} (cyan atoms indicate Cs; purple atoms indicate I; black atoms indicate Pb).}
\end{figure}

Crystalline materials often exist in multiple structural forms, called phases. For example, ice shows a remarkable degree of polymorphism\cite{Bore2023, Reinhardt2021QuantumWater, Salzmann2019AdvancesWaterPhaseDiagram}. Despite water's abundance and apparent simplicity, its strong dipole moment and bend geometry cause water molecules to organize in complex hydrogen-bonded networks. As a result, ice exhibits over 20 different crystalline phases, with even more theoretically predicted\cite{Hansen2021Everlasting}. The computational determination of the most stable phase of a given solid at certain conditions is a key aspect of materials design. For instance, \ce{CsPbI3}, a metal halide perovskite (MHP) depicted in Fig.~\ref{fig:materials}, is investigated for its use in solar cells, but only the so-called black phase exhibits favorable opto-electronic properties. Unfortunately, the black phase is only stable at high temperature and quickly converts to the inactive yellow phase at ambient conditions\cite{Steele2019Thermal}. The development of physico-chemical techniques to stabilize the black phase at low temperatures is a very active field of research\cite{Braeckevelt2022Accurately}. Thermodynamically, the most stable phase at a given pressure and temperature is the one with the lowest Gibbs free energy, making accurate Gibbs free energy calculations crucial for computational material design\cite{tuckerman2010statistical}.

\begin{figure}
    \centering
    \includegraphics[width=1.0\linewidth]{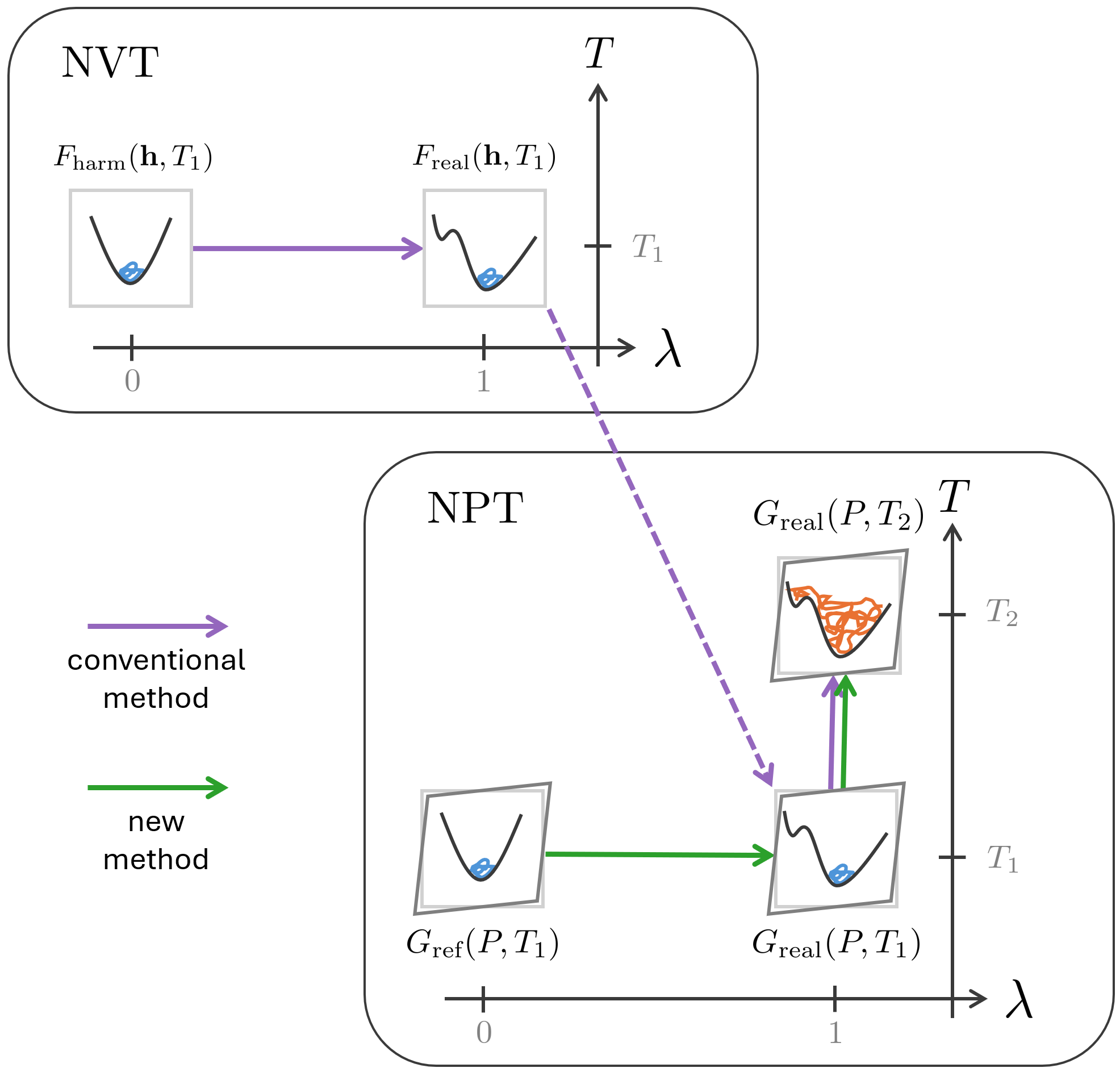}
    \caption{\label{fig:main_intro} \textbf{Schemes for computing Gibbs free energies via thermodynamic integration (TI). Purple arrows indicate the conventional scheme,} in which the NVT harmonic reference is first corrected towards the true potential energy surface (PES). A second \textit{approximate} correction is performed from isochoric to isobaric conditions and, finally, a third from low to high temperature\cite{Cheng2018}. \textbf{Green arrows indicate the new scheme,} where the reference is a newly derived NPT harmonic PES. Subsequent corrections account for anharmonicity and temperature effects, all performed under constant-pressure conditions.}
\end{figure}

Nowadays, Thermodynamic Integration (TI) is the standard approach for obtaining accurate Gibbs free energies of a phase, as it provides a versatile framework for exact free energy calculations under realistic conditions\cite{Kapil2019Assessment, Cheng2018, Kapil2022}. The state-of-the-art TI scheme is described by Cheng and Ceriotti\cite{Cheng2018}, and is illustrated by the purple arrows in Fig.~\ref{fig:main_intro}. The procedure starts from a reference free energy, and requires 3 successive corrections. The reference value is the Helmholtz free energy of the NVT harmonic approximation of the crystal, which yields an analytical expression as a function of its vibrational frequencies. The first correction of the conventional TI scheme accounts for anharmonic effects by integrating from the harmonic reference to the real crystal at fixed cell and temperature. The second correction incorporates volume flexibility, thereby converting the Helmholtz free energy into Gibbs free energy. Importantly, this NVT-to-NPT correction is approximate, as outlined hereafter. Finally, a third correction accounts for temperature effects.

A limitation of the conventional TI scheme is that the conversion from Helmholtz to Gibbs free energy is not fully rigorous. While the conventional correction relies on a volume distribution obtained via constant-pressure simulations, the exact correction requires the full cell-shape distribution of the material, which is computationally very expensive. While the volume-approximation is reasonably suited for rigid or single-minimum crystals, accuracy issues might arise for flexible materials with multiple (cell) minima, because the volume distribution will not be able to capture the underlying complexity of the six-dimensional multimodal cell-shape distribution. This limitation motivated the development of the new method presented below.

We introduce a new TI scheme that enables the computation of exact Gibbs free energy differences without approximations and without increase in computational cost relative to the conventional approach. The scheme is illustrated by the green arrows in Fig.~\ref{fig:main_intro} and operates fully within the NPT ensemble, thereby avoiding the NVT-to-NPT conversion. Crucially, we introduce a novel NPT reference crystal, which is approximately the harmonic expansion of the real crystal at constant pressure. The main theoretical contribution of this work is the derivation of the NPT reference potential and its Gibbs free energy expression. The Gibbs free energy of the real crystal at any pressure and temperature is obtained by applying only two corrections to the reference value: (i) an anharmonic correction, computed via TI from the harmonic to real crystal at constant pressure and temperature, and (ii) a temperature correction identical to the final step in the conventional method.

To assess the performance and accuracy of the proposed method, we consider two complementary case studies, depicted in Fig.~\ref{fig:materials}. The first case study examines three ice phases selected for their simple (i.e., unimodal) cell-shape distribution. The second focuses on \ce{CsPbI3}, whose black phase exhibits degenerate cell minima\cite{PhysRevMaterials, Bechtel2019FiniteTemperaturePerovskites}, yielding a complicated (i.e., multimodal) cell-shape distribution. For the former system, we demonstrate that the new method reproduces results obtained with the conventional approach, while for the latter, we observe systematic deviations between the two methods, which can be attributed to inaccuracies introduced by the volume-approximation of the NVT-to-NPT correction. Moreover, the novel method requires comparable computational cost and offers improved transparency of the workflow.

The remainder of this article is organized as follows. The \emph{Methodology} section reviews the conventional framework and presents the derivation of the new reference Gibbs free energy. The \emph{Computational details} section describes the new thermodynamic integration workflow and summarizes the simulation protocols employed for both case studies. Finally, the \emph{Results and discussion} section analyzes the two case studies, focusing on the accuracy and practical applicability of the proposed method in comparison to the conventional approach.

\section{Methodology}

This section is divided into two parts. First, we review the conventional TI scheme described by Cheng and Ceriotti\cite{Cheng2018}. Second, we present a detailed derivation of the newly proposed NPT harmonic reference, which constitutes the core methodological advance of this work.

\subsection{Conventional TI scheme}

The conventional TI scheme begins from a reference Helmholtz free energy and applies three successive corrections to obtain the exact Gibbs free energy at the thermodynamic conditions of interest. This workflow is schematically illustrated by the purple arrows in Fig.~\ref{fig:main_intro}.

\paragraph{Reference.}
The reference quantity is the Helmholtz free energy of the crystal within the NVT harmonic approximation. Its analytical expression at temperature $T_1$ and fixed simulation cell $\mathbf{h}$ reads:
\begin{equation}
    \label{eq:Fref}
    F_{\text{harm}}(\mathbf{h}, T_1) = k_\text{B}T_1 \sum_{i=1}^{3N-3} \ln \left(\frac{\hbar\omega_i}{k_\text{B}T_1}\right) + E_\text{opt}(\mathbf{h})
\end{equation}
where $k_\text{B}$ is the Boltzmann constant, $\hbar$ is the reduced Planck constant, $E_\text{opt}(\mathbf{h})$ denotes the optimized crystal energy for cell $\mathbf{h}$, and $\omega_i$ are the $3N-3$ non-zero harmonic vibrational frequencies. The translational contribution to the free energy associated with the three zero modes is omitted, as it is negligible for a macroscopic crystal. We report the derivation of Eq.~\ref{eq:Fref} in Section 1 of the Supporting Information, as it is instructive for our new methodology.

\paragraph{Anharmonic correction.}
The first correction is computed via TI from the harmonic approximation to the real crystal, at fixed cell $\mathbf{h}$ and temperature $T_1$:
\begin{equation}
    \label{eq:conv_corr_1}
        F_{\text{real}}(\mathbf{h},T_1) = 
        F_{\text{harm}}(\mathbf{h},T_1) + \int_0^1 d\lambda \left\langle U_{\text{real}} - U_{\text{harm}} \right\rangle_{\lambda, \mathbf{h}, T_1}
\end{equation}
Here, $U_{\text{harm}}$ and $U_{\text{real}}$ denote the potential energy surface (PES) of the harmonic and real crystal, respectively. The coupling parameter $\lambda$ linearly interpolates between the two: $U(\lambda) = (1-\lambda)U_\text{harm} + \lambda U_\text{real}$, thereby defining the TI path.

\paragraph{NVT-to-NPT correction.}
The second correction incorporates volume flexibility yielding an approximate value for the Gibbs free energy of the real crystal at temperature $T_1$ and pressure $P$:
\begin{equation}
    \label{eq:convFtoG}
G_{\text{real}}(P,T_1) \approx F_{\text{real}}(\mathbf{h}, T_1) + P V + k_\text{B} T_1 \ln \left( \rho(V \mid P, T_1) \right)
\end{equation}
where $V = \det(\mathbf{h})$ is the cell volume and $\rho(V \mid P, T_1)$ is the probability of observing volume $V$ at pressure $P$ and temperature $T_1$. This one-dimensional distribution can be obtained by accumulating a volume histogram during NPT MD simulations. Importantly, Eq.~\ref{eq:convFtoG} is approximate, as the exact correction would require the full six-dimensional cell-shape distribution $\rho(\mathbf{h} \mid P, T_1)$, which is computationally very expensive to compute.

\paragraph{Temperature correction.}
Finally, a third correction is applied via a second TI step by integrating the temperature from $T_1$ to $T_2$:
\begin{equation}
    \label{eq:tempcorrNPT}
\frac{G_{\text{real}}(T_2)}{k_\text{B} T_2} = \frac{G_{\text{real}}(T_1)}{k_\text{B} T_1} - \int_{T_1}^{T_2} \frac{\langle U + K + PV \rangle_{P, T}}{k_\text{B} T^2} \, dT
\end{equation}
where $U$ and $K$ are the potential and kinetic energy, respectively. Alternatively, this temperature correction may be applied in the NVT ensemble, prior to the NVT-to-NPT correction. In that case, the ensemble average $\langle U + K + PV \rangle_{P, T}$ in Eq.~\ref{eq:tempcorrNPT} is replaced by $\langle U + K \rangle_{\mathbf{h}, T}$.

\subsection{Derivation of the novel NPT harmonic reference}

The proposed TI scheme, illustrated by the green arrows in Fig.~\ref{fig:main_intro}, comprises a reference Gibbs free energy followed by two TI corrections. The second TI correction is given by Eq.~\ref{eq:tempcorrNPT} and is the same as in the conventional scheme, while the first one reads:
\begin{equation}
    \label{eq:GTI}
    G_{\text{real}}(P,T_1) = 
        G_{\text{ref}}(P,T_1) + \int_0^1 d\lambda \left\langle U_{\text{real}} - U_{\text{ref}} \right\rangle_{\lambda, P, T_1}
\end{equation}
The key methodological contribution of this work is the derivation of a suitable reference Gibbs free energy, $G_{\text{ref}}$. The central challenge lies in defining an appropriate reference potential $U_{\text{ref}}$ with flexible cell degrees of freedom. Crucially, $U_{\text{ref}}$ must satisfy the two following criteria: (i) it must yield an analytical expression for its Gibbs free energy, and (ii) it must facilitate rapid convergence of the TI correction in Eq.~\ref{eq:GTI}. The construction of $U_{\text{ref}}$ and the corresponding derivation of $G_{\text{ref}}$ are developed in detail in the following sections.

\subsubsection{Initial setup}

The starting point for the derivation of $G_\text{ref}$ (and $U_\text{ref}$) is the classical expression for the anisotropic NPT partition function\cite{tuckerman2010statistical}:
\begin{equation}
    \label{eq:startingpoint}
        Q_\text{ref}(P,T) =\frac{1}{V_s} \int_{0}^{\infty} \mathrm{d}\mathbf{h} \, \frac{1}{V^2} e^{-\beta P V} \\ \frac{1}{h^{3N}} \int_{-\infty}^{\infty} \mathrm{d}\mathbf{p}_i^{N} e^{-\beta K } \int_{0}^{\mathbf{h}} \mathrm{d}\mathbf{r}_i^{N} e^{-\beta U_\text{ref}}
\end{equation}
where $V = \det(\mathbf{h})$ is the cell volume and $\mathbf{h}$ is the 3x3 matrix containing the 3 cell vectors. $\beta$ is $1/k_\text{B}T$, $h$ is the Planck constant, and $V_s$ is a normalization volume. The integrandum contains three Boltzmann factors: one for the $PV$ term, one for the kinetic energy $K$, and one for the potential energy $U$ of the system. All three Boltzmann factors are accompanied by a matching integration: over 9 cell components of $\mathbf{h}$, over the momenta $\mathbf{p}_i$ of all N particles, and over the positions $\mathbf{r}_i$ of all N particles, respectively.

The left-hand-side can be rewritten using the relation between the Gibbs free energy and the NPT partition function:
\begin{equation}
    \label{eq:GiskTlnD}
    G_\text{ref}(P,T)=-k_\text{B}T\ln(Q_\text{ref}(P,T))
\end{equation}
The right-hand-side can be simplified by integrating over the atomic momenta:
\begin{equation}
    \label{eq:momenta}
    \frac{1}{h^{3N}} \int_{-\infty}^{\infty} \mathrm{d}\mathbf{p}_i^{N} e^{-\beta K} = \prod_{i=1}^{N} \left(\frac{2 \pi m_i}{\beta h^2}\right)^{3/2} = \prod_{i=1}^{N} \left(\frac{1}{\lambda_i}\right)^3
\end{equation}
where $\lambda_i$ is the thermal wavelength of atom $i$. Combining equations \ref{eq:startingpoint}, \ref{eq:GiskTlnD}, \ref{eq:momenta} yields:
\begin{equation}
    \label{eq:endstartingpoint}
    e^{-\beta G_\text{ref}(P,T)} = \prod_{j=1}^{N} \left(\frac{1}{\lambda_j}\right)^3 \frac{1}{V_s} \int_{0}^{\infty} \mathrm{d}\mathbf{h} \, \frac{1}{V^2} e^{-\beta P V} \int_{0}^{\mathbf{h}} \mathrm{d}\mathbf{r}_i^{N} e^{-\beta U_\text{ref}}
\end{equation}

\subsubsection{Transformation to deformed coordinates}

The combination of a fluctuating cell and Periodic Boundary Conditions (PBC) during NPT simulations can cause unphysical behavior when Cartesian coordinates are used. Namely, the system's energy will depend on the arbitrary chosen boundaries of the simulation cell, as the Cartesian positions lack coupling to the cell. Instead, scaled or deformed coordinates are suited for NPT crystalline systems\cite{Alavi2020}. Scaled - also called fractional - coordinates are dimensionless and express atomic positions relative to the lattice vectors of the cell. Here, we opt for deformed coordinates as their unit is length, consistent with the cell degrees of freedom, such that all matrix elements of the extended Hessian (vide infra) have the same dimension (namely [energy/length$^2$]). Deformed coordinates are defined as:
\begin{equation}
    \label{eq:deformedcoo}
    \mathbf{d}_i = \mathbf{r}_i \cdot \mathbf{h}^{-1} \cdot  \mathbf{h}_0
\end{equation}
where we use the row-major convention. $\mathbf{r}_i = \begin{bmatrix}r_{i,x} & r_{i,y} & r_{i,z} \end{bmatrix}$ is the Cartesian position of atom $i$, $\mathbf{d}_i = \begin{bmatrix}d_{i,x} & d_{i,y} & d_{i,z} \end{bmatrix}$ is its deformed position, $\mathbf{h}$ is the current cell, and $\mathbf{h}_0$ is a yet-to-be-defined reference cell. The deformed coordinates of an atom automatically adapt to changes in the cell-shape and size. In other words, the atomic positions relative to the unit cell will not change if the latter is subjected to deformations.

The Jacobian matrix for the transformation of one atom is $\mathbf{J} = (\mathbf{h}^{-1} \cdot \mathbf{h}_0)^T$. As we do this transformation for N atoms, the following factor must be added to the integrandum: $\frac{1}{\det(\mathbf{J})^N} = \frac{V^N}{V_0^N}$. Applying the transformation from Cartesian coordinates to deformed coordinates on expression \ref{eq:endstartingpoint} yields:
\begin{equation}
    \label{eq:trans}
        e^{-\beta G_\text{ref}(P,T)} = \prod_{j=1}^{N} \left( \frac{1}{\lambda_j} \right)^3 \frac{1}{V_s V_0^N}
    \int_{0}^{\infty} \mathrm{d}\mathbf{h}\, e^{-\beta P V} V^{N-2}
    \int_{0}^{\mathbf{h}_0} \mathrm{d}\mathbf{d}_i^{N} e^{-\beta U_\text{ref}}
\end{equation}

\subsubsection{NPT harmonic approximation}
We rewrite the volume factor:
\begin{equation}
    \label{eq:V_is_e_ln_V}
    V^{N-2} = e^{(N-2)\ln(V)}
\end{equation}
where we note that taking the natural logarithm of the volume does not yield any problems regarding units, because changing units would only yield an additional constant term to the final free energy, which cancels out when comparing different phases. Applying Eq.~\ref{eq:V_is_e_ln_V} on Eq.~\ref{eq:trans} yields the following integrandum:
\begin{equation}
    \label{eq:allexp}
        e^{-\beta \left(U_\text{ref} + P V -  \frac{N-2}{\beta} \ln(V)\right)}=e^{-\beta \left(U_\text{ref} + U_\text{bias}\right)}
\end{equation}
where the newly defined bias potential $U_\text{bias} = P V -  \frac{N-2}{\beta} \ln(V)$ depends on the system size, temperature, and pressure.

Up to this point, the derivation is generic for any reference potential $U_\text{ref}$. However, hereafter, we will derive an explicit expression for $U_\text{ref}$ based on the two criteria mentioned in the beginning of the derivation: $U_\text{ref}$ should yield an analytically integrable $G_\text{ref}$ and should facilitate TI convergence. These constraints give raise to the definition of $U_\text{ref}$.

\textbf{Constraint 1:} We choose the exponent of Eq.~\ref{eq:allexp} to be quadratic, such that the integrandum has a Gaussian-like expression, allowing analytical integration. Therefore, $U_\text{ref}$ is defined as $\mathcal{T}^{\,2}(U_f) - U_\text{bias}$. The integrandum becomes $e^{-\beta \mathcal{T}^{\,2}(U_f)}$, where $\mathcal{T}^{\,2}(U_f)$ denotes the second-order Taylor expansion (w.r.t. atomic and cell coordinates) of the function $U_f$ around its minimum.

\textbf{Constraint 2:} For rapid convergence of $\langle U_{\text{real}} - U_{\text{ref}} \rangle_{\lambda,P,T}$, the reference PES should closely approximate the real PES. To achieve this, we define: $U_f = U_{\text{real}} + U_\text{bias}$. This results in the final expression for $U_\text{ref}$:
\begin{equation}
    \label{eq:Uref}
    U_\text{ref} = \mathcal{T}^{\,2}(U_{\text{real}} + U_\text{bias}) - U_\text{bias}
\end{equation}
where $U_\text{bias}$ is included in the Taylor expansion to compensate the subsequent subtraction. Hence, the reference potential is approximately the NPT harmonic expansion of the real potential. We note that for a constant cell, $U_f$ and $U_{\text{real}}$ are identical up to a constant offset because $U_\text{bias}$ equals $P V -  \frac{N-2}{\beta} \ln(V)$.

Since $U_\text{ref}$ is now defined, we proceed with the derivation of its Gibbs free energy $G_\text{ref}$. Therefore, we insert our new definition of $U_\text{ref}$ in Eq.~\ref{eq:trans}:
\begin{equation}
    \label{eq:intermediate}
    e^{-\beta G_\text{ref}(P,T)}
    = \prod_{j=1}^{N} \left( \frac{1}{\lambda_j} \right)^3 \frac{1}{V_s V_0^N}
    \int_{0}^{\infty} \mathrm{d}\mathbf{h} \int_{0}^{\mathbf{h}_0} \mathrm{d}\mathbf{d}_i^{N} e^{-\beta \mathcal{T}^{\,2}(U_{\text{real}} + U_\text{bias})}
\end{equation}
where we will write $U_{\text{real}} + U_\text{bias} = U_f$ for the remainder of the derivation. The second-order Taylor expansion $\mathcal{T}^{\,2}(U_f)$, derived hereafter, is constructed around its minimum $(\mathbf{d}_{i,0}, \mathbf{h}_0)$. This guarantees that the exponent of the integrandum has a trivial constant term and a quadratic term, but no linear term. Also, we now formally define the coordinate transformation of Eq.~\ref{eq:deformedcoo}, namely the optimized cell $\mathbf{h}_0$ of $U_f$ is a suitable choice as reference cell.

When all $3N$ deformed coordinates and all 9 cell coordinates are put in a single row $\mathbf{x}$ of length $3N+9$, then $\mathcal{T}^{\,2}(U_f)$ can be compactly written as:
\begin{equation}
    \label{eq:t2energy}
    \mathcal{T}^{\,2}(U_f)= U_{f,0} + \frac{1}{2} [ \mathbf{x} - \mathbf{x}_0 ] \cdot \mathbf{H}_{\mathrm{ext}} \cdot [ \mathbf{x} - \mathbf{x}_0 ]^T
\end{equation}
where $U_{f,0}$ is the energy at the minimum of $U_f$, and $\mathbf{x}_0$ contains the optimized geometry in a row of length $3N+9$, using the same format as $\mathbf{x}$. $\mathbf{H}_{\mathrm{ext}}$ is the extended Hessian, i.e. a $3N+9$ by $3N+9$ matrix describing the curvature of $U_f$ at the minimum $(\mathbf{d}_{i,0}, \mathbf{h}_0)$. It is computed by taking second-order derivatives with respect to deformed coordinates and cell coordinates and has the following block-matrix form:
\begin{equation}
    \mathbf{H}_{\mathrm{ext}} = 
    \begin{bmatrix}
    \frac{\partial^2 U_f}{\partial \mathbf{d} \partial \mathbf{d}} & \frac{\partial^2 U_f}{\partial \mathbf{d} \partial \mathbf{h}} \\
    \frac{\partial^2 U_f}{\partial \mathbf{h} \partial \mathbf{d}} & \frac{\partial^2 U_f}{\partial \mathbf{h} \partial \mathbf{h}} \\
    \end{bmatrix}
\end{equation}

Inserting the second order Taylor expansion from Eq.~\ref{eq:t2energy} in Eq.~\ref{eq:intermediate} gives:
\begin{equation}
    \label{eq:Hinexp}
    e^{-\beta G_\text{ref}(P,T)}
    = \prod_{j=1}^{N} \left( \frac{1}{\lambda_j} \right)^3 \frac{1}{V_s V_0^N}
    \int_{0}^{\infty} \mathrm{d}\mathbf{h} \int_{0}^{\mathbf{h}_0} \mathrm{d}\mathbf{d}_i^{N} e^{-\beta \left(U_{f,0} + \frac{1}{2} [ \mathbf{x} - \mathbf{x}_0 ] \cdot \mathbf{H}_{\mathrm{ext}} \cdot [ \mathbf{x} - \mathbf{x}_0 ]^T\right)}
\end{equation}

\subsubsection{Diagonalization}
As $\mathbf{H}_{\mathrm{ext}}$ is not a diagonal matrix, $[ \mathbf{x} - \mathbf{x}_0 ] \cdot \mathbf{H}_{\mathrm{ext}} \cdot [ \mathbf{x} - \mathbf{x}_0 ]^T$ still contains undesirable mixed terms. Therefore, the next step is to diagonalize the extended Hessian: $\mathbf{H}_{\mathrm{ext}} = \mathbf{N}^T \cdot \mathbf{D} \cdot \mathbf{N}$, with $\mathbf{D} = \mathrm{diag}(D_1, D_2, \dots, D_{3N+9})$, and $\mathbf{N}$ the matrix containing the eigenvectors of the system. As $\mathbf{H}_{\mathrm{ext}}$ is a symmetric matrix, $\mathbf{N}$ is orthogonal.

A second and final coordinate transformation is performed to coordinates $q_i$:
\begin{equation}
    \label{eq:transdia}
    \mathbf{q} = [ \mathbf{x} - \mathbf{x}_0 ] \cdot \mathbf{N}^T
\end{equation}
For this transformation, the Jacobian determinant is equal to one. The $3N+9$ coordinates $q_i$ contain 3 translational modes, 3 rotational modes (crystal rotations are allowed as $\mathbf{h}$ has 9 cell components), and $3N+3$ vibrational modes. After applying the transformation of Eq.~\ref{eq:transdia} to Eq.~\ref{eq:Hinexp}, we find independent Gaussian integrals for each vibrational mode $q_i$:
\begin{equation}
    \label{eq:almostthere}
    e^{-\beta G_\text{ref}(P,T)} = \prod_{j=1}^{N} \left( \frac{1}{\lambda_j} \right)^3 \frac{e^{-\beta U_{f,0}}}{V_s V_0^N} 
    \int_{-\infty}^{+\infty} \mathrm{d} q_i^{3N+3} \, e^{- \frac{\beta}{2} \sum_{i=1}^{3N+3} D_i q_i^2}
\end{equation}
where the right-hand-side is not multiplied with the translational nor the rotational partition function, as their contribution to the free energy of a macroscopic crystal is negligible. The integral boundaries of the vibrational modes are approximated to extend from $-\infty$ to $+\infty$ as this allows analytical integration, and will only add a negligible contribution to the free energy because the integrandum decreases with $e^{-q^2}$ when diverging from the minimum.

\subsubsection{Final expression}
The $3N+3$ decoupled Gaussian integrals in Eq.~\ref{eq:almostthere} are solved via $\int_{-\infty}^{+\infty}e^{-ax^2} dx= \sqrt{\frac{\pi}{a}}$, yielding:
\begin{equation}
    e^{-\beta G_\text{ref}(P,T)} = \prod_{j=1}^{N} \left( \frac{1}{\lambda_j} \right)^3 \frac{e^{-\beta U_{f,0}}}{V_s V_0^N} 
    \prod_{i=1}^{3N+3} \sqrt{\frac{2\pi}{\beta D_i}}
\end{equation}
Finally, the analytical expression for the reference Gibbs free energy is obtained:
\begin{equation}
    \label{eq:finalgref}
    G_\text{ref}(P,T) =  U_{f,0}-\frac{1}{\beta} \ln \left( \frac{1}{V_s V_0^N} \right)
    - \frac{3}{\beta} \ln \left( \prod_{i=1}^{N} \frac{1}{\lambda_i} \right)
    - \frac{1}{2\beta} \ln \left( \prod_{i=1}^{3N+3} \frac{2\pi}{\beta D_i} \right)
\end{equation}
where $U_{f,0}$ is the optimized energy of the function $U_f = U_{\text{real}} + P V -  \frac{N-2}{\beta} \ln(V)$ and $V_0$ the volume of the optimized cell. $V_s$ is a normalization volume, $\lambda_i$ is the thermal wavelength of atom $i$, and $D_i$ is a non-zero eigenvalue of the extended Hessian of $U_f$. Overall, this new derivation yields an analytical expression for the reference Gibbs free energy $G_\text{ref}$, forming the theoretical cornerstone of the novel NPT TI scheme.

\section{Computational details}

Herein, we discuss the workflow of the new TI scheme, and summarize the simulation protocols employed for the ice and \ce{CsPbI3} case study.

\subsection{Workflow of the new TI scheme}

\begin{figure}
    \centering
    \includegraphics[width=1.0\linewidth]{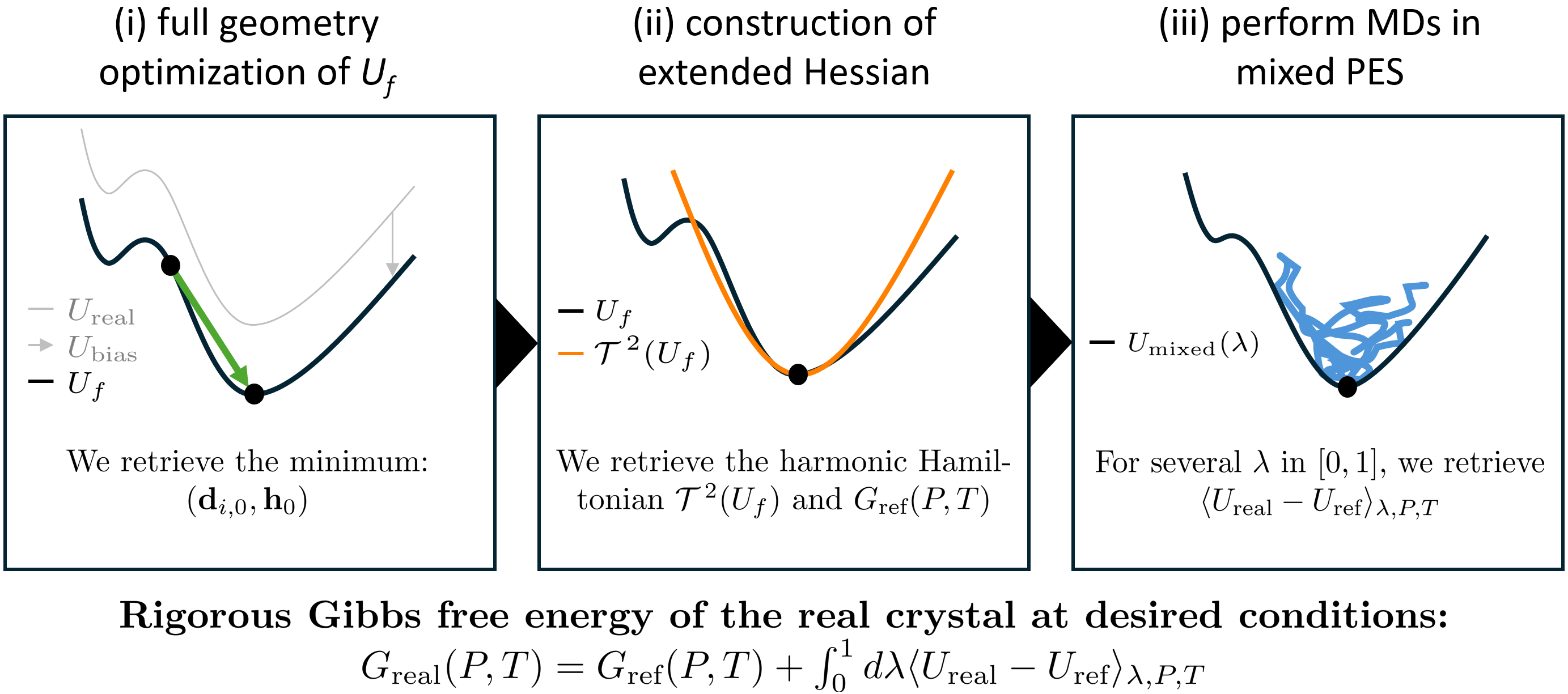}
    \caption{\label{fig:implementation_visual_overview}\textbf{Overview of the computational workflow of the new TI scheme.} (i) A full geometry optimization yields the optimized atomic positions and simulation cell. (ii) At this minimum, the extended Hessian is constructed via second-order derivatives of deformed and cell coordinates. The extended Hessian defines the NPT harmonic PES, and its non-zero eigenvalues yield the reference Gibbs free energy. (iii) For several $\lambda$-values in the interval $[0,1]$ an NPT MD simulation is run in the mixed PES (Eq.~\ref{eq:mixedHam}). The $\lambda$-TI correction is found via trapezoidal integration. Overall, two inexpensive steps yield the reference Gibbs free energy, while the computationally more demanding TI-step corrects towards the real value.}
\end{figure}

Having established the theoretical framework for the reference Gibbs free energy $G_{\text{ref}}$, we now describe the computational procedure used to obtain the true Gibbs free energy $G_{\text{real}}$. The workflow consists of three main steps, which are graphically illustrated in Fig.~\ref{fig:implementation_visual_overview}. A full implementation\cite{zenodo18632301,THERMOFLOW_git} is provided in Python, on top of our in-house developed molecular simulation library Psiflow\cite{vandenhaute2023mlmof}, which makes use of ASE, PLUMED, MACE, and i-PI\cite{Larsen2017ASE, plumed2019, batatia2023macehigherorderequivariant, iPI}.

The first step is a full geometry optimization of the function $U_f = U_{\text{real}} + U_{\text{bias}}$, yielding the optimized geometry $(\mathbf{d}_{i,0}, \mathbf{h}_0)$ and the corresponding minimum energy $U_{f,0}$. The optimization enforces vanishing atomic forces and zero stress on the simulation cell. Due to the specific form of the bias term $U_\text{bias} = P V -  \frac{N-2}{\beta} \ln(V)$, high pressure and low temperature favor small cells, whereas low pressure and high temperature favor larger cells.

The second step is the construction of the extended Hessian $\mathbf{H}_{\mathrm{ext}}$, after which Eq.~\ref{eq:finalgref} can be evaluated directly to obtain the reference Gibbs free energy. In this work, $\mathbf{H}_{\mathrm{ext}}$ is assembled row by row. Analytical first-order derivatives of the energy are obtained using a Machine Learning Interatomic Potential (MLIP), and second-order derivatives are obtained via finite-differences. A detailed description of the extended Hessian construction protocol is provided in Section 2 of the Supporting Information.

The third step is to evaluate the TI correction, which requires molecular dynamics (MD) simulations in the mixed PES:
\begin{equation}
    \label{eq:mixedHam}
U_\text{mixed}(\lambda) = U_\text{ref} +  \lambda(U_\text{real}-U_\text{ref}) = (1 - \lambda)\left( \mathcal{T}^{\,2}(U_f) - U_\text{bias} \right) + \lambda U_{\text{real}}
\end{equation}
The central technical challenge is enabling MD simulations in the NPT harmonic PES $\mathcal{T}^{\,2}(U_f)$, which requires energy, forces, and stresses for any input geometry. Their derivation and implementation is explained in Section 2 of the Supporting Information.

\subsection{Case studies}

In this section, we provide a qualitative overview of the used TI-schemes of both case studies. For technical specifications (i.e. thermostat/barostat type\cite{langevin1908brownian, Martyna10041996_MTTK}, number of replicas, $\lambda$-values, etc.), we refer to Section 3 of the Supporting Information.

\subsubsection{Ice}
The three investigated ice phases are ice II, IX, and XI, as all three are proton-ordered phases and have been experimentally observed at different conditions (see Fig.~\ref{fig:materials})\cite{Hansen2021Everlasting}. We derive Gibbs free energies for all three phases at eight thermodynamic conditions, namely at 0.1 / 1000 MPa, and at 30 / 60 / 120 / 240 K. The ice phases have a simulation cell with 96 water molecules and are simulated with a Machine Learning Interatomic Potential (MLIP), using the MACE architecture\cite{batatia2023macehigherorderequivariant, Batatia2022mace, batatia2025design}. The model was trained from scratch using the revPBE-D3(BJ) level of theory (details on MLIP generation are reported in Section 4 of the Supporting Information)\cite{Zhang1998revPBE, Grimme2011D3BJ, grimme2010consistent}. As the TI protocol requires long MD runs, the use of a MLIP is pivotal in making these simulations feasible. The ice phases are chosen as case study because their cell has a single minimum during NPT simulations and randomly fluctuates, resulting in a simple cell-shape distribution with a single peak. 

\begin{figure}
    \centering
    \includegraphics[width=0.85\linewidth]{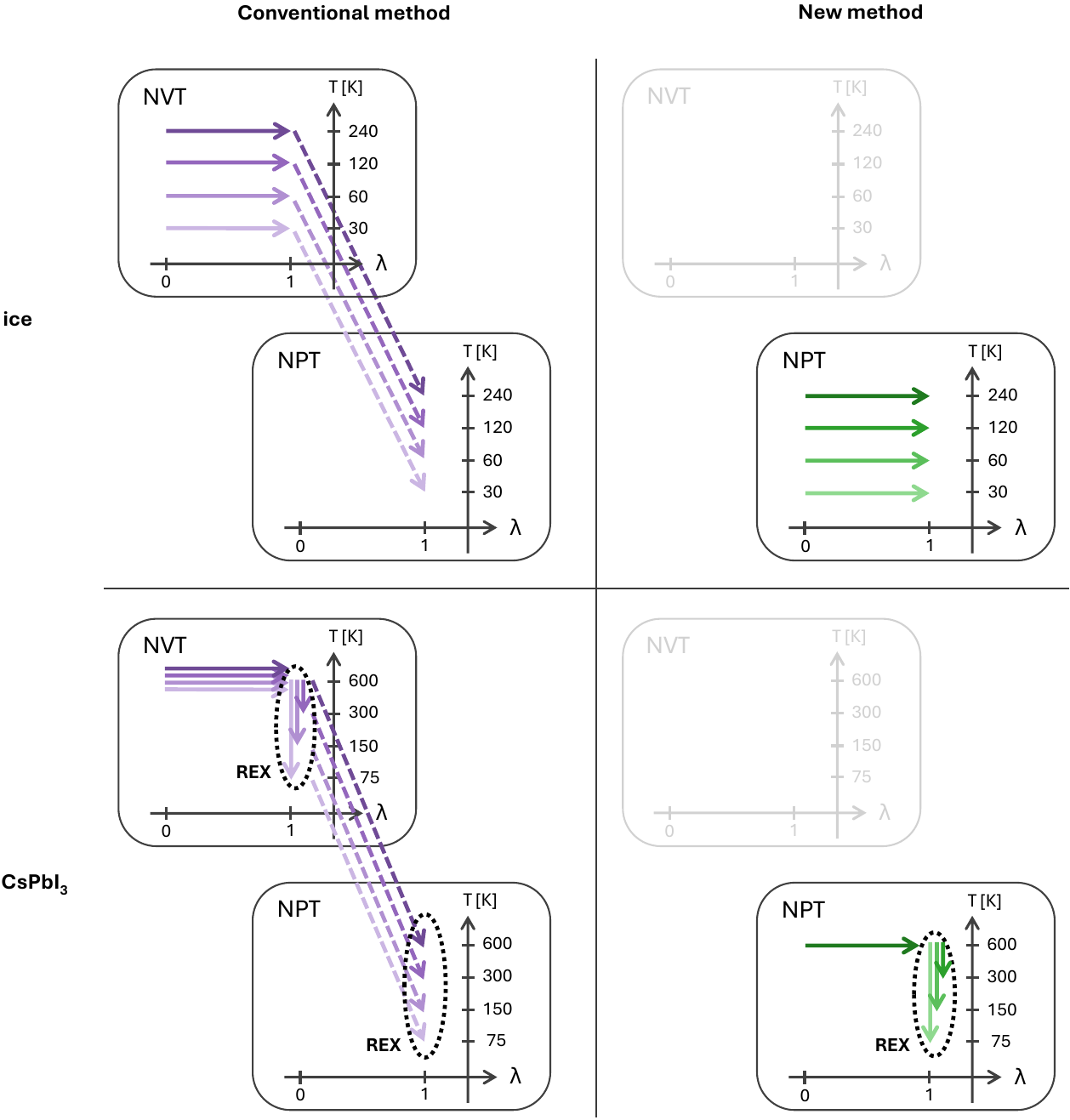}
    \caption{\label{fig:big_scheme}\textbf{Overview of the applied TI schemes.} \textbf{Ice-conventional:} NVT reference and both corrections are computed at the final temperature. \textbf{Ice-new:} NPT reference and single correction are computed at the final temperature and pressure. \textbf{\ce{CsPbI3}-conventional:} Each desired temperature yields an average cell, yielding four references and four NVT $\lambda$-TI corrections, computed at 600 K to properly sample all tilted configurations of the \ce{PbI6} octahedra. NVT temperature corrections are applied using Replica Exchange (REX) MD to ensure proper sampling at low temperature\cite{Earl2005ParallelTempering}. The NVT-to-NPT correction is applied at four different temperatures, via histograms accumulated during NPT REX MD. \textbf{\ce{CsPbI3}-new:} A single reference at 600 K is computed, followed by a single NPT $\lambda$-TI correction. For the lower temperatures, NPT temperature corrections are applied using REX.}
\end{figure}

In Fig.~\ref{fig:big_scheme}, the upper half illustrates the TI schemes for the ice case study, which are detailed hereafter. For the new method, the TI procedure is straightforward: (i) the reference Gibbs free energy is computed at the desired temperature and pressure, and (ii) the NPT $\lambda$-TI is performed at the same conditions.

For the conventional method, the TI procedure is more complex. First of all, two steps instead of one are required, as the NVT-to-NPT correction $PV + k_\text{B} T \ln \left(\rho(V \mid P, T)\right)$ comes into play. Also, we should be aware that $\rho(V \mid P, T)$ is prone to noise when $V$ is sampled poorly for the given $P$ and $T$. Therefore, the ideal situation is that the NVT simulations are performed at a volume that maximizes $\rho(V \mid P, T)$, as suggested by Cheng and Ceriotti\cite{Cheng2018}. This means that one \textit{first} does the NPT simulations, computes the most probable volume, and only then does NVT $\lambda$-TI at that volume. The problem with this approach is that there is no one-to-one relation between the volume $V$ and the cell $\mathbf{h}$, because multiple cells will satisfy $\det(\mathbf{h}) = V$. To avoid this ambiguity in our workflow, we computed the average cell during NPT simulations and used this cell for the NVT $\lambda$-TI. This preliminary NPT simulation step is often referred to as the NPT equilibration step\cite{Li2025AbInitioMelting, ChewReinhardt2023}. With previous considerations in mind, the following procedure was used for the conventional method: (i) run NPT simulations at desired pressure and temperature, and compute the average cell, (ii) compute the reference Helmholtz free energy at the same conditions, and (iii) run NVT $\lambda$-TI at the same conditions, and compute the NVT-to-NPT correction via the simulations of (i).

\subsubsection{\ce{CsPbI3}}
For \ce{CsPbI3}, we investigate the black and yellow phase (see Fig.~\ref{fig:materials}). While the black phase has a perovskite structure and is thermodynamically favorable at elevated temperatures ($>$ 550 K), the yellow phase is stable at room temperature and below\cite{Braeckevelt2022Accurately}. We derive the Gibbs free energies for the black and yellow phase at atmospheric pressure and at four temperatures (75 / 150 / 300 / 600 K). The \ce{CsPbI3} simulation cell contains eight units (i.e. 40 atoms) and simulations were performed with a MLIP using the MACE architecture\cite{batatia2023macehigherorderequivariant, Batatia2022mace, batatia2025design}. The \textit{ab initio} data was generated via PBE+D3(BJ) level of theory (details on MLIP generation are reported in Section 4 of the Supporting Information)\cite{Perdew1996PBE, Grimme2011D3BJ, grimme2010consistent}. The \ce{CsPbI3} case study was chosen as its black phase has a complicated cell-shape distribution. This originates from the fact that the cell has not one but six (degenerate) cell minima, characterized by the cell angles: two out of three angles are 90°, the third one is either 87° or 93°, as is visualized in Section 5 of the Supporting Information. From an atomistic point of view, these cell minima originate from multiple tilted configurations of the \ce{PbI6} octahedra\cite{PhysRevMaterials, Bechtel2019FiniteTemperaturePerovskites}. Due to these minima, the direct calculation of $G_{\text{real}}$ from $G_{\text{ref}}$ used for ice, is not applicable for \ce{CsPbI3}. Spontaneous transitions between these tilted configurations occur at 600 K, but not at 300 K or lower. Therefore, in order to properly sample all minima, the computationally demanding $\lambda$-TI simulations were performed at 600 K. The subsequent temperature correction is computed via MD simulations at multiple temperatures, coupled via Replica Exchange (REX; also known as parallel tempering), assuring proper sampling at lower temperatures\cite{Earl2005ParallelTempering, OKABE2001435, SUGITA1999141}. As was the case for ice, the MLIP is crucial to make these intensive MD simulations computationally feasible. For the conventional and new method, the TI-scheme is depicted in Fig.~\ref{fig:big_scheme}.

For the new method, we adopted the following procedure: (i) the reference Gibbs free energy is computed at 600 K and atmospheric pressure, (ii) the NPT $\lambda$-TI correction is computed at the same temperature and pressure, and (iii) three temperature corrections are computed via NPT REX MD simulations\cite{OKABE2001435}.

For the conventional method, the procedure is more complex: (i) run REX NPT MD simulations at all four temperatures and atmospheric pressure, and compute the average cell for each $(P,T)$-condition, (ii) compute the reference Helmholtz free energy for each of the four cells at 600 K, (iii) run NVT $\lambda$-TI at 600 K for each of the four cells, (iv) compute a temperature correction via NVT REX MDs\cite{SUGITA1999141}, and compute the NVT-to-NPT correction via the NPT simulations of (i).

\section{Results and discussion}
In this section, we present and discuss the Gibbs free energies obtained via the conventional and new method, applied on ice and \ce{CsPbI3} phases. We end with a discussion comparing the computational cost and usability of both methods.

\subsection{Ice}
\begin{figure}
    \centering
    \includegraphics[width=1.0\linewidth]{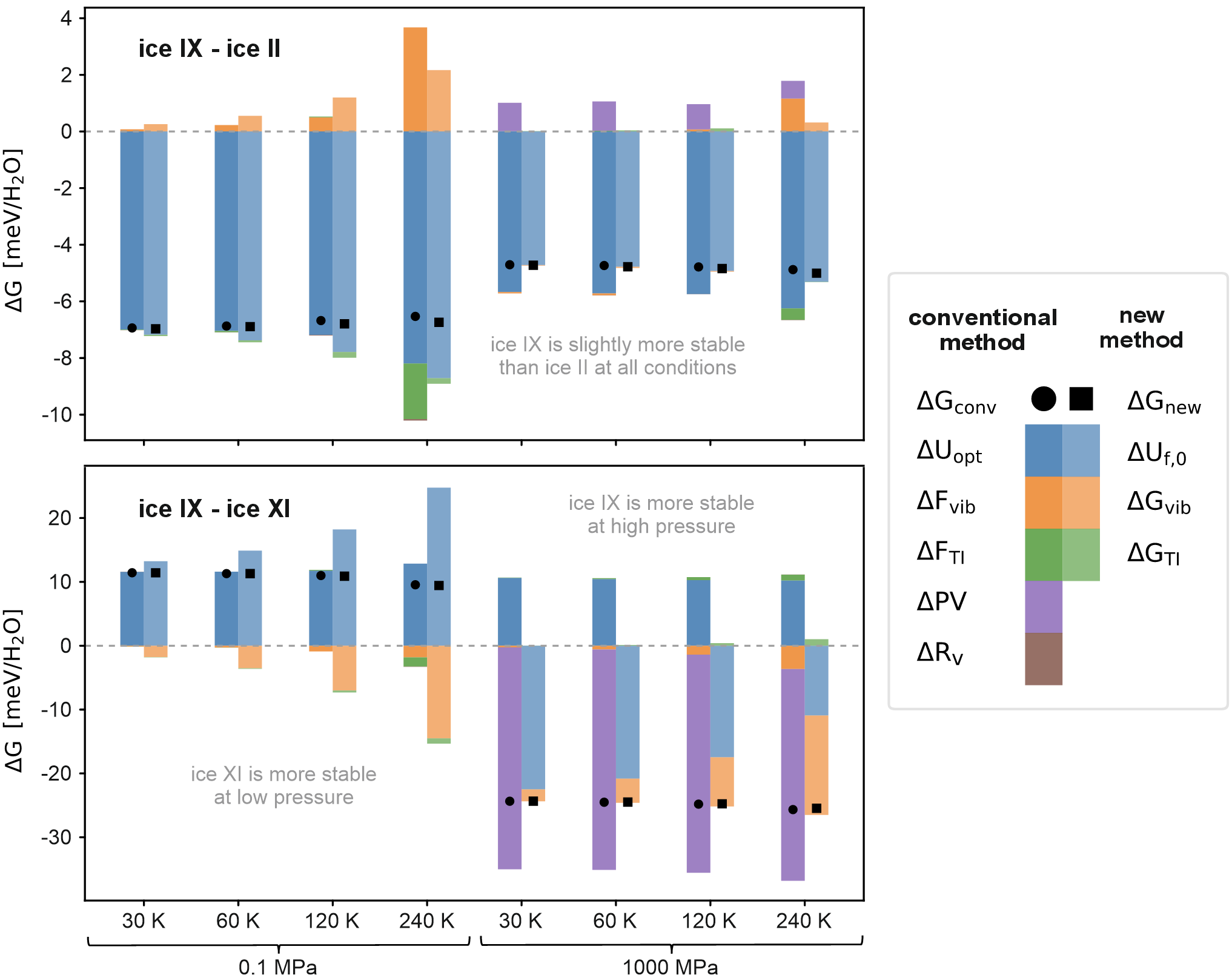}
    \caption{\label{fig:result_ice}\textbf{Gibbs free energy differences and their contributions for ice.} Results obtained via the conventional method and the new method, at various pressures and temperatures. The conventional method has five free energy contributions, whereas the new method has only three. Positive contributions are stacked above the zero axis and negative contributions below. In all cases, the total Gibbs free energy difference $\Delta G_\text{conv}$ and $\Delta G_\text{new}$ agree extremely well, as expected for materials with simple cell-shape distributions.}
\end{figure}

In this first case study, we demonstrate that the new method reproduces the Gibbs free energy differences of the conventional method when the investigated material has a simple cell-shape distribution.

Fig.~\ref{fig:result_ice} shows the Gibbs free energy differences computed via both methods, at multiple temperatures and pressures, and for two phase-combinations: ice IX versus ice II, and ice IX versus ice XI. Crucially, the total Gibbs free energy differences predicted by the two approaches are in excellent agreement. The largest deviation between the methods is 0.337 meV per $\text{H}_2\text{O}$ molecule, while all estimated errors remain below 0.1 meV per $\text{H}_2\text{O}$ molecule.

The colored bars in Fig.~\ref{fig:result_ice} represent the individual contributions to $\Delta G$ and highlight three important aspects. (i) The different thermodynamic conditions lead to substantially varying free energy contributions, demonstrating that both methods are probed in physically distinct regimes. (ii) The total $\Delta G$ of each method is the result of a complex interplay of the different contributions, indicating that none of the corrections can be neglected. (iii) The contribution types and their magnitudes are unique for each method, emphasizing that the new method is not a variation on the conventional method, but really a completely new approach, with a fundamentally different theoretical framework. Despite these (i) diverse thermodynamic conditions, (ii) intertwined contributions, and (iii) distinct approaches, we find an excellent agreement between both methods, which validates the theoretical foundations and practical execution of our new method.

A closer inspection of the individual free energy contributions in Fig.~\ref{fig:result_ice} provides additional physical insight into both approaches.

The conventional method decomposes the Gibbs free energy difference into five terms:
\begin{equation}
    \Delta G_\text{conv} = \Delta U_\text{opt} + \Delta F_\text{vib} + \Delta F_\text{TI} + \Delta PV + \Delta R_V
\end{equation}
where $U_{opt}$ is the optimized energy of the system at volume $V$, and $F_\text{vib}$ is the vibrational free energy at fixed volume, obtained from the eigenvalues of the mass-weighted Hessian. Their sum constitutes the harmonic Helmholtz free energy (see Eq.~\ref{eq:Fref}). The anharmonic correction computed via NVT $\lambda$-TI is the contribution $F_\text{TI}$. The final two terms account for the NVT-to-NPT correction, where $R_V = k_\text{B} T \ln \left( \rho(V \mid P, T) \right)$.

In contrast, the new method consists of only three contributions:
\begin{equation}
    \Delta G_\text{new} = \Delta U_{f,0} + \Delta G_\text{vib} + \Delta G_\text{TI}
\end{equation}
where $U_{f,0}$ is the optimized energy of $U_f$, and $G_\text{vib}$ is the vibrational free energy at constant pressure, computed from the eigenvalues of the extended Hessian. Together, $U_{f,0}$ and $G_\text{vib}$ constitute the reference Gibbs free energy (see Eq.~\ref{eq:finalgref}; note that $G_\text{vib}$ also contains $k_\text{B}T \ln \left( V_s V_0^N \right)$). The third contribution $G_\text{TI}$ is the anharmonic correction computed via NPT $\lambda$-TI.

Interpreting the results of Fig.~\ref{fig:result_ice} through the thermodynamic identity $\Delta G = \Delta E - T\Delta S + P\Delta V$ reveals trends that are consistent across both methods. At low temperatures and low pressures, $\Delta G$ is dominated by the ground-state energy difference, such that $\Delta G \approx \Delta E$. At elevated pressures, the $P\Delta V$ term becomes increasingly important, yielding $\Delta G \approx \Delta E + P\Delta V$. While the $P\Delta V$ contribution is negligible at atmospheric pressure, it plays a significant role at 1~GPa, particularly for the ice IX versus ice XI comparison due to the larger molar volume of ice XI (30.6~\AA$^3$ per \ce{H2O}) compared to ice IX (24.7~\AA$^3$ per \ce{H2O}). In the conventional method, this contribution appears explicitly as a separate term, whereas in the new method it is implicitly included in $U_{f,0}$ via the definition of $U_f$. At low pressures and high temperatures, vibrational and anharmonic contributions become increasingly important, reflecting the growing role of entropy, such that $\Delta G \approx \Delta E - T\Delta S$. Across all conditions, the $\Delta R_V$ contribution is negligible, which can be attributed to the similar and well-behaved volume distributions of the three ice phases, leading to the cancellation of this term.

Finally, we compare our results with experimental phase stability. Ice XI is experimentally known to be more stable than ice IX at atmospheric pressure and less stable at high pressure, in agreement with our calculations\cite{Hansen2021Everlasting}. While we correctly predict ice IX to be more stable than ice II at low temperatures, we do not find ice II to be stable at 240~K. Given the small free energy differences involved, this discrepancy is consistent with the expected absolute errors at the revPBE level of theory\cite{Santra2011IcePRL}.

\subsection{\ce{CsPbI3}}

In this second case study, we demonstrate that the new method can outperform the conventional method, due to the pronounced discrepancy between the volume and cell-shape distribution of the black phase of \ce{CsPbI3}. Therefore, we begin by analyzing and comparing these distributions for the black and yellow phase.

The origin of this discussion traces back to the NVT-to-NPT correction of the conventional method (Eq.~\ref{eq:convFtoG}). We define:
\begin{equation}
    R_V = k_\text{B} T \ln (\rho(V \mid P,T)), \quad
    R_{\mathbf{h}} = k_\text{B} T \ln (\rho(\mathbf{h} \mid P,T))
\end{equation}
where $R_V$ is the conventionally used expression, and $R_{\mathbf{h}}$ is its rigorous (but computationally intractable) form. The inaccuracy of the conventional method for $\Delta G_\text{conv}$ can be expressed as the difference between the rigorous $\Delta R_{\mathbf{h}}$ and the approximate $\Delta R_V$:
\begin{equation}
    \label{eq:DDR}
    \Delta R_{\mathbf{h}} - \Delta R_V =
    k_\text{B} T
    \ln \left(
    \frac{\rho(\mathbf{h}_\text{black} \mid P,T)}
         {\rho(\mathbf{h}_\text{yellow} \mid P,T)}
    \cdot
    \frac{\rho(V_\text{yellow} \mid P,T)}
         {\rho(V_\text{black} \mid P,T)}
    \right),
\end{equation}
where $\mathbf{h}$ denotes the simulation cell used in the NVT calculations, and $V$ is its volume. Eq.~\ref{eq:DDR} shows that the inaccuracy of the conventional method scales linearly with temperature, and scales logarithmically with the discrepancy between the probability of the average cell and the average volume of a phase. Also, while other free energy contributions like the vibrational ones, scale with system size, the correction of Eq.~\ref{eq:DDR} does not, such that the inaccuracy becomes less relevant for larger systems.

\begin{figure}
    \centering
    \includegraphics[width=1.0\linewidth]{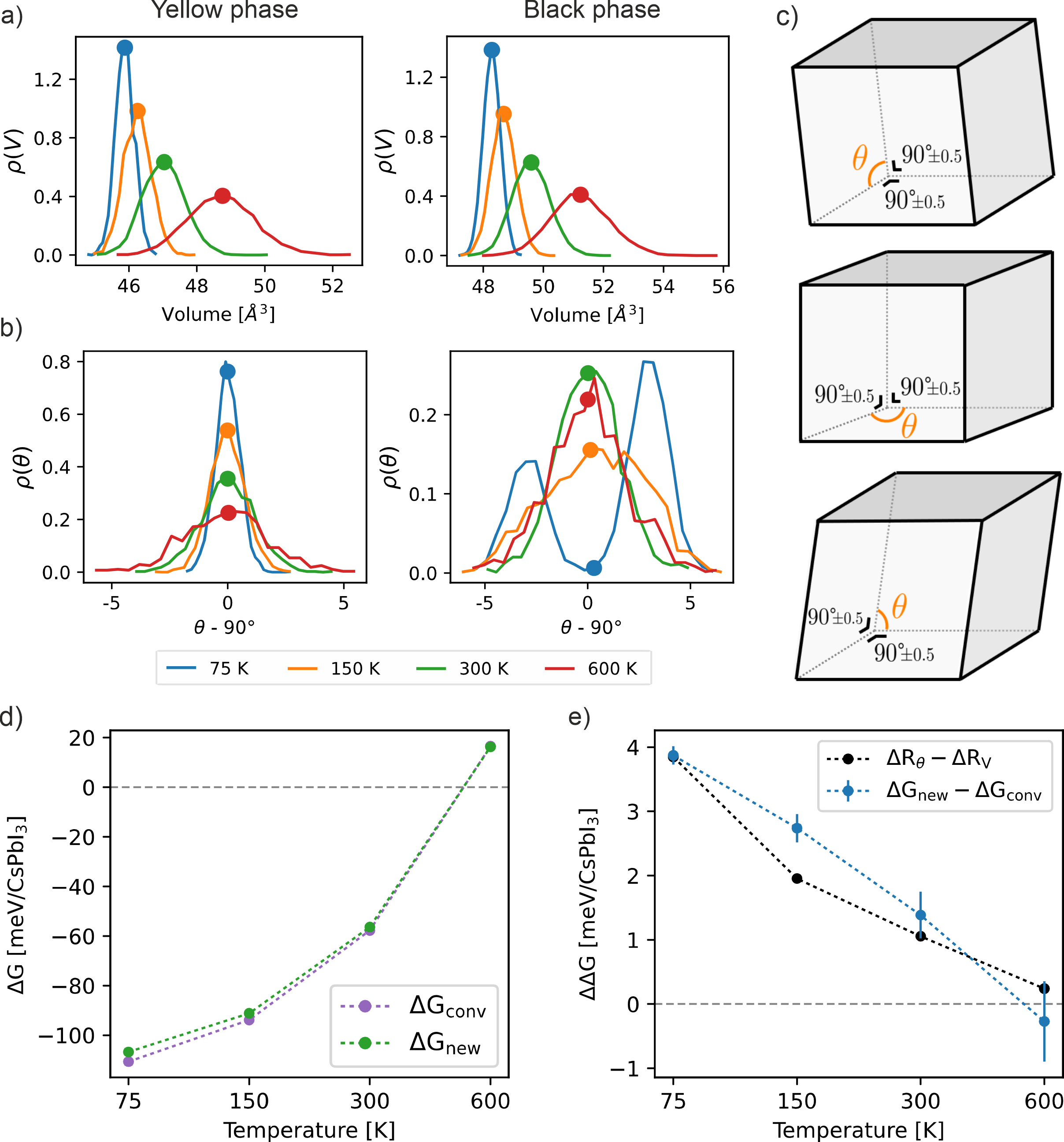}
    \caption{\label{fig:result_cspbi3}\textbf{Overview of the \ce{CsPbI3} results.} 
    (a) Volume distributions of the yellow and black phase.
    (b) $\theta$-distributions of the yellow and black phase.
    (c) Schematic illustration of our \textit{ad hoc} one-dimensional variable $\theta$.
    (d) Gibbs free energy differences between yellow and black phase.
    (e) Difference of $\Delta G$ between the two methods and an estimate of the inaccuracy of the conventional method.}
\end{figure}

As shown in Fig.~\ref{fig:result_cspbi3}a, both phases of \ce{CsPbI3} exhibit similar and normal volume distributions across all temperatures considered. The probability of the average volume is nearly identical for the black and yellow phase, implying that the correction $\Delta R_V$ is negligibly small at all temperatures.

In contrast, the two phases differ strongly in their cell-shape distributions. The black phase exhibits six degenerate shallow minima characterized by the cell angles $(\alpha,\beta,\gamma)$:
\[
(90^\circ,90^\circ,87^\circ),\ (90^\circ,90^\circ,93^\circ),\ 
(90^\circ,87^\circ,90^\circ),\ (90^\circ,93^\circ,90^\circ),\ 
(87^\circ,90^\circ,90^\circ),\ (93^\circ,90^\circ,90^\circ),
\]
whereas the yellow phase possesses a single minimum at $(90^\circ,90^\circ,90^\circ)$. In principle, assessing the inaccuracy of the conventional method via Eq.~\ref{eq:DDR} requires knowledge of the full cell-shape distribution $\rho(\mathbf{h})$. However, computing this distribution is precisely what the approximate $\rho(V)$ correction was designed to avoid. We therefore introduce an \textit{ad hoc} one-dimensional metric, denoted $\theta$, which captures the essential features of the cell-shape fluctuations more effectively than the volume.

The construction of $\theta$ is motivated by the observation that, at low temperatures, the black phase fluctuates around one of its six distinct cell minima, while the average cubic cell with angles $(90^\circ,90^\circ,90^\circ)$ is highly improbable. Our goal is to quantify the unlikeliness of this average cell by estimating the probability that all three cell angles are simultaneously close to $90^\circ$. To this end, $\theta$ is defined as the \textit{third angle} of the simulation cell, conditional on the other two angles being close to $90^\circ$ (see Fig.~\ref{fig:result_cspbi3}c). Concretely, we impose a cutoff of $0.5^\circ$ on two angles and record the third as a $\theta$ value. This procedure is repeated cyclically for all three angle combinations. A three-dimensional scatter plot of $(\alpha,\beta,\gamma)$ at 75~K and 600~K, together with the angle-selection procedure, is shown in SI~6.

The resulting $\theta$-distributions are shown in Fig.~\ref{fig:result_cspbi3}b. For the yellow phase, the distributions are approximately normally distributed, with decreasing width at lower temperatures. In contrast, for the black phase at 75~K, the distribution exhibits two pronounced maxima at $90^\circ \pm 3^\circ$, while the average angle of $90^\circ$ is strongly suppressed. At higher temperatures, the distributions broaden significantly. These results confirm that the black and yellow phases have fundamentally different cell-shape distributions, which should give rise to distinct NVT-to-NPT corrections. However, this effect is entirely missed by the volume-based correction $\rho(V)$, whereas our \textit{ad hoc} one-dimensional metric $\theta$ captures it explicitly. As such, $\theta$ provides a practical estimator for the inaccuracy of the conventional method.

Figure~\ref{fig:result_cspbi3}d shows the Gibbs free energy difference between the yellow and black phases computed using both methods. The yellow phase is predicted to be more stable at 75, 150, and 300~K, while the black phase becomes more stable at 600~K, in agreement with experimental observations\cite{Steele2019Thermal, Braeckevelt2022Accurately}. Although the two methods yield similar overall trends, Fig.~\ref{fig:result_cspbi3}e reveals systematic deviations at lower temperatures.

Given the rigor of our theoretical framework and the known approximation underlying the conventional method, we attribute this discrepancy to the term $\Delta R_{\mathbf{h}} - \Delta R_V$ in Eq.~\ref{eq:DDR}. We estimate this inaccuracy using the newly introduced variable $\theta$:
\begin{equation}
    \Delta R_{\mathbf{h}} - \Delta R_V \approx \Delta R_\theta - \Delta R_V.
\end{equation}
The resulting estimate, shown as the black curve in Fig.~\ref{fig:result_cspbi3}e, closely matches the observed difference between the two methods. This suggests that the new method is indeed more accurate, as it fully accounts for complicated cell-shape fluctuations. Nevertheless, the absolute gain in accuracy remains modest, implying that the conventional method is sufficient in many practical applications.

\subsection{Comparing computational cost and usability}

Beyond accuracy, computational cost and practical convenience are important considerations. The ice case study allows for a straightforward comparison. For the conventional method, we performed NVT MD simulations at ten $\lambda$ values, each consisting of $10^5$ steps, together with a single NPT simulation of $10^5$ steps to compute the NVT-to-NPT correction. This amounts to a total of $11 \times 10^5$ steps per free energy computation (i.e., per pressure, per temperature, and per ice phase). For the new method, we performed NPT simulations at ten $\lambda$ values, again with $10^5$ steps each, yielding a total of $10 \times 10^5$ steps per free energy computation. Consequently, the new method requires approximately 9\% fewer computational resources. On the other hand, the estimated free energy error of the conventional method is on average about 5\% smaller. These two effects compensate, such that the two methods can be considered practically equal in computational cost. Regarding the absolute cost, the computationally intensive TI simulations were made feasible by MLIPs, allowing over 10 MD steps per second, such that all ice calculations required less than 1400 GPU hours. In Section 3.2 of the Supporting Information, an analysis of the computational expense for the \ce{CsPbI3} case study is provided, affirming that both methods have a analogous computational cost.

Finally, we assess the usability of both methods. As illustrated in Fig.~\ref{fig:big_scheme}, the TI workflow of the conventional method is more complex. It involves an additional NVT-to-NPT correction step and requires the selection of a simulation cell for the NVT calculations, a choice that is not uniquely defined and may affect reproducibility and accuracy. In contrast, the new method follows a transparent and well-defined two-step procedure. These two considerations make clear that the new method excels as a user-friendly method.

\section{Conclusion}
Thermodynamic Integration (TI) is widely regarded as the state-of-the-art computational framework for obtaining accurate Gibbs free energies of crystalline solids. The conventional TI methodology follows a 3-step scheme starting from a harmonic reference in the NVT ensemble. A major limitation of this approach is the NVT-to-NPT correction, which requires a computationally intractable six-dimensional cell-shape distribution. In the literature, the NVT-to-NPT correction is approximated via a one-dimensional volume distribution, thereby neglecting full cell flexibility.

In this work, we introduce a new TI framework that operates entirely in the NPT ensemble, eliminating the need for the approximate NVT-to-NPT correction. The new reference is a harmonic system with flexible cell degrees of freedom, yielding an analytical expression for its reference Gibbs free energy. Starting from this reference, the Gibbs free energy of the real system at arbitrary thermodynamic conditions is obtained through only two subsequent corrections: an anharmonic correction and a temperature correction.

The accuracy of the proposed NPT TI framework was assessed using two complementary case studies. For ice polymorphs, the Gibbs free energy differences obtained via both methods match excellently, demonstrating that the new method faithfully reproduces the conventional results for systems with simple cell-shape distributions. In contrast, for \ce{CsPbI3}, whose black phase exhibits a complicated cell-shape distribution, both methods yield slightly deviating Gibbs free energy differences, particularly at lower temperatures. Via an \textit{ad hoc} one-dimensional cell-shape metric, we estimated the inaccuracy introduced by the volume-based approximation in the conventional method. This estimated error closely matches the observed deviations between the two approaches, indicating that the new method predicts more accurate Gibbs free energies for complex materials. Nevertheless, the magnitude of the correction remains modest, indicating that the conventional method remains adequate for many applications.

Beyond accuracy considerations, the new framework offers practical advantages. From a computational perspective, its overall cost is practically equal to that of the conventional approach. More importantly, by operating entirely within the NPT ensemble, the new method significantly simplifies the workflow for Gibbs free energy calculations, replacing the conventional three-step procedure with a more transparent and conceptually straightforward two-step scheme.

In summary, we introduced a novel NPT thermodynamic integration scheme that maintains comparable computational cost while offering improved theoretical rigor and a simplified workflow.

\begin{acknowledgement}
M.B. thanks the Research Foundation—Flanders (FWO) for a junior postdoctoral fellowship (grant n. 1269725N). V.V.S. acknowledges funding from the Research Board of Ghent University, from the European Research Council (ERC) under the European Union's Horizon 2020 research and innovation programme 101201153 (ERC-AdG-2024, TIME-project), and from iBOF-21-085 PERsist. The resources and services used in this work were provided by the VSC (Flemish Supercomputer Center), funded by the Research Foundation - Flanders (FWO) and the Flemish Government. Additionally, we acknowledge the EuroHPC Joint Undertaking for awarding this project access to the EuroHPC supercomputer LUMI, hosted by CSC (Finland) and the LUMI consortium.

\end{acknowledgement}

\begin{suppinfo}

Input files, relevant datasets, and analysis scripts used to obtain the results presented in this manuscript are available on Zenodo (https://doi.org/10.5281/zenodo.18632301). Any additional data is available from the authors upon request.\\
SI~1: Helmholtz free energy of the NVT harmonic approximation of a crystal\\
SI~2: Implementation of the novel NPT TI scheme\\
SI~3: Computational details\\
SI~4: MLIP generation\\
SI~5: Visualization of tilted cell configurations of the black phase of \ce{CsPbI3}\\
SI~6: Interactive 3D scatter plot of $(\alpha,\beta,\gamma)$ at 75~K and 600~K for black and yellow phase of \ce{CsPbI3}
\end{suppinfo}

\bibliography{biblio}

@article{Hansen2021Everlasting,
  author  = {Hansen, Thomas C.},
  title   = {The everlasting hunt for new ice phases},
  journal = {Nature Communications},
  year    = {2021},
  volume  = {12},
  pages   = {3161},
  doi     = {10.1038/s41467-021-23403-6}
}

@article{Steele2019Thermal,
  author  = {Steele, J. A. and Jin, H. and Dovgaliuk, I. and Berger, R. F. and Braeckevelt, T. and Yuan, H. and Martin, C. and Solano, E. and Lejaeghere, K. and Rogge, S. M. J. and Notebaert, C. and Vandezande, W. and Janssen, K. P. F. and Goderis, B. and Debroye, E. and Wang, Y.-K. and Dong, Y. and Ma, D. and Saidaminov, M. and Tan, H. and Lu, Z. and Dyadkin, V. and Chernyshov, D. and Van Speybroeck, V. and Sargent, E. H. and Hofkens, J. and Roeffaers, M. B. J.},
  title   = {Thermal unequilibrium of strained black {CsPbI}\textsubscript{3} thin films},
  journal = {Science},
  year    = {2019},
  volume  = {365},
  number  = {6454},
  pages   = {679--684},
  doi     = {10.1126/science.aax3878}
}

@article{Braeckevelt2022Accurately,
  author  = {Braeckevelt, Tom and Goeminne, Ruben and Vandenhaute, Sander and Borgmans, Sander and Verstraelen, Toon and Steele, Julian A. and Roeffaers, Maarten B. J. and Hofkens, Johan and Rogge, Sven M. J. and Van Speybroeck, Veronique},
  title   = {Accurately determining the phase transition temperature of {CsPbI}\textsubscript{3} via random-phase approximation calculations and phase-transferable machine learning potentials},
  journal = {Chemistry of Materials},
  year    = {2022},
  volume  = {34},
  number  = {19},
  pages   = {8561--8576},
  doi     = {10.1021/acs.chemmater.2c01508}
}

@book{tuckerman2010statistical,
  author    = {Tuckerman, Mark E.},
  title     = {Statistical Mechanics: Theory and Molecular Simulation},
  edition   = {2},
  publisher = {Oxford University Press},
  address   = {Oxford},
  year      = {2010},
  isbn      = {9780198525264}
}

@article{Cheng2018,
  author  = {Cheng, Bingqing and Ceriotti, Michele},
  title   = {Computing the absolute Gibbs free energy in atomistic simulations: Applications to defects in solids},
  journal = {Physical Review B},
  year    = {2018},
  volume  = {97},
  number  = {5}
}

@article{Kapil2022,
  author  = {Kapil, Venkat and Schran, Christoph and Zen, Andrea and Chen, Ji and Pickard, Chris J. and Michaelides, Angelos},
  title   = {The first-principles phase diagram of monolayer nanoconfined water},
  journal = {Nature},
  year    = {2022},
  volume  = {609},
  number  = {7927},
  pages   = {512--516}
}

@article{Kapil2019Assessment,
  author  = {Kapil, Venkat and Engel, Edgar and Rossi, Mariana and Ceriotti, Michele},
  title   = {Assessment of approximate methods for anharmonic free energies},
  journal = {Journal of Chemical Theory and Computation},
  year    = {2019},
  volume  = {15},
  number  = {11},
  pages   = {5845--5857},
  doi     = {10.1021/acs.jctc.9b00596}
}

@article{vandenhaute2023mlmof,
  author  = {Vandenhaute, S. and Cools-Ceuppens, M. and DeKeyser, S. and Verstraelen, T.},
  title   = {Machine learning potentials for metal-organic frameworks using an incremental learning approach},
  journal = {npj Computational Materials},
  year    = {2023},
  volume  = {9},
  pages   = {19}
}

@article{plumed2019,
  author  = {{PLUMED Consortium}},
  title   = {Promoting transparency and reproducibility in enhanced molecular simulations},
  journal = {Nature Methods},
  year    = {2019},
  volume  = {16},
  pages   = {670--673}
}

@article{iPI,
  author  = {Litman, Yair and Kapil, Venkat and Feldman, Yotam M. Y. and Tisi, Davide and Begu{\v{s}}i{\'c}, Tomislav and Fidanyan, Karen and Fraux, Guillaume and Higer, Jacob and Kellner, Matthias and Li, Tao E. and P{\'o}s, Eszter S. and Stocco, Elia and Trenins, George and Hirshberg, Barak and Rossi, Mariana and Ceriotti, Michele},
  title   = {i-{PI} 3.0: A flexible and efficient framework for advanced atomistic simulations},
  journal = {Journal of Chemical Physics},
  year    = {2024},
  volume  = {161},
  number  = {6},
  pages   = {062504}
}

@article{Larsen2017ASE,
  author  = {Larsen, Ask H. and Mortensen, Jens J{\o}rgen and Blomqvist, Jakob and others},
  title   = {The atomic simulation environment --- a Python library for working with atoms},
  journal = {Journal of Physics: Condensed Matter},
  year    = {2017},
  volume  = {29},
  number  = {27},
  pages   = {273002},
  doi     = {10.1088/1361-648X/aa680e}
}

@misc{zenodo18632301,
  author    = {De Witte, K.},
  title     = {NPT TI SCHEME},
  year      = {2026},
  publisher = {Zenodo},
  doi       = {10.5281/zenodo.18632301},
  url       = {https://doi.org/10.5281/zenodo.18632301}
}

@misc{THERMOFLOW_git,
  author = {Braeckevelt, Tom},
  title  = {ThermoFlow},
  year   = {2025},
  note   = {GitHub repository. Available at \url{https://github.com/tbraeckevelt/thermoflow}}
}

@article{Zhang1998revPBE,
  author  = {Zhang, Yan and Yang, Weitao},
  title   = {Comment on ``generalized gradient approximation made simple''},
  journal = {Physical Review Letters},
  year    = {1998},
  volume  = {80},
  number  = {4},
  pages   = {890--890},
  doi     = {10.1103/PhysRevLett.80.890}
}

@article{Perdew1996PBE,
  author  = {Perdew, John P. and Burke, Kieron and Ernzerhof, Matthias},
  title   = {Generalized gradient approximation made simple},
  journal = {Physical Review Letters},
  year    = {1996},
  volume  = {77},
  number  = {18},
  pages   = {3865--3868},
  doi     = {10.1103/PhysRevLett.77.3865}
}

@article{Grimme2011D3BJ,
  author  = {Grimme, Stefan and Ehrlich, Stephan and Goerigk, Lars},
  title   = {Effect of the damping function in dispersion corrected density functional theory},
  journal = {Journal of Computational Chemistry},
  year    = {2011},
  volume  = {32},
  number  = {7},
  pages   = {1456--1465},
  doi     = {10.1002/jcc.21759}
}

@article{grimme2010consistent,
  author  = {Grimme, Stefan and Antony, Jens and Ehrlich, Stephan and Krieg, Helge},
  title   = {A consistent and accurate ab initio parametrization of density functional dispersion correction ({DFT-D}) for the 94 elements {H--Pu}},
  journal = {Journal of Chemical Physics},
  year    = {2010},
  volume  = {132},
  number  = {15},
  doi     = {10.1063/1.3382344}
}

@article{PhysRevMaterials,
  author  = {Bechtel, Jonathon S. and Van der Ven, Anton},
  title   = {Octahedral tilting instabilities in inorganic halide perovskites},
  journal = {Physical Review Materials},
  year    = {2018},
  volume  = {2},
  number  = {2},
  pages   = {025401},
  doi     = {10.1103/PhysRevMaterials.2.025401}
}

@article{SUGITA1999141,
  author  = {Sugita, Yuji and Okamoto, Yuko},
  title   = {Replica-exchange molecular dynamics method for protein folding},
  journal = {Chemical Physics Letters},
  year    = {1999},
  volume  = {314},
  number  = {1},
  pages   = {141--151},
  doi     = {10.1016/S0009-2614(99)01123-9}
}

@article{OKABE2001435,
  author  = {Okabe, Tsuneyasu and Kawata, Masaaki and Okamoto, Yuko and Mikami, Masuhiro},
  title   = {Replica-exchange {Monte Carlo} method for the isobaric-isothermal ensemble},
  journal = {Chemical Physics Letters},
  year    = {2001},
  volume  = {335},
  number  = {5},
  pages   = {435--439},
  doi     = {10.1016/S0009-2614(01)00055-0}
}

@article{Earl2005ParallelTempering,
  author  = {Earl, David J. and Deem, Michael W.},
  title   = {Parallel tempering: Theory, applications, and new perspectives},
  journal = {Physical Chemistry Chemical Physics},
  year    = {2005},
  volume  = {7},
  number  = {23},
  pages   = {3910--3916},
  doi     = {10.1039/B509983H}
}

@book{Alavi2020,
  author    = {Alavi, Saman},
  title     = {Molecular Simulations: Fundamentals and Practice},
  publisher = {Wiley-VCH},
  address   = {Weinheim},
  year      = {2020}
}

@article{Bore2023,
  author  = {Bore, S. L. and Paesani, F.},
  title   = {Realistic phase diagram of water from ``first principles'' data-driven quantum simulations},
  journal = {Nature Communications},
  year    = {2023},
  volume  = {14},
  number  = {1},
  pages   = {3349},
  doi     = {10.1038/s41467-023-38855-1}
}

@article{Reinhardt2021QuantumWater,
  author  = {Reinhardt, Aleks and Cheng, Bingqing},
  title   = {Quantum-mechanical exploration of the phase diagram of water},
  journal = {Nature Communications},
  year    = {2021},
  volume  = {12},
  number  = {1},
  pages   = {588},
  doi     = {10.1038/s41467-020-20821-w}
}

@article{Salzmann2019AdvancesWaterPhaseDiagram,
  author  = {Salzmann, Christoph G.},
  title   = {Advances in the experimental exploration of water's phase diagram},
  journal = {Journal of Chemical Physics},
  year    = {2019},
  volume  = {150},
  number  = {6},
  pages   = {060901},
  doi     = {10.1063/1.5085163}
}

@article{Bechtel2019FiniteTemperaturePerovskites,
  author  = {Bechtel, Jonathon S. and Thomas, John C. and Van der Ven, Anton},
  title   = {Finite-temperature simulation of anharmonicity and octahedral tilting transitions in halide perovskites},
  journal = {Physical Review Materials},
  year    = {2019},
  volume  = {3},
  number  = {11},
  pages   = {113605},
  doi     = {10.1103/PhysRevMaterials.3.113605}
}

@article{Martyna10041996_MTTK,
  author  = {Martyna, Glenn J. and Tuckerman, Mark E. and Tobias, Douglas J. and Klein, Michael L.},
  title   = {Explicit reversible integrators for extended systems dynamics},
  journal = {Molecular Physics},
  year    = {1996},
  volume  = {87},
  number  = {5},
  pages   = {1117--1157},
  doi     = {10.1080/00268979600100761}
}

@article{langevin1908brownian,
  author  = {Langevin, Paul},
  title   = {Sur la th{\'e}orie du mouvement brownien},
  journal = {Comptes Rendus de l'Acad{\'e}mie des Sciences},
  year    = {1908},
  volume  = {146},
  pages   = {530--533}
}

@article{Santra2011IcePRL,
  author  = {Santra, Biswajit and Michaelides, Angelos and Fuchs, Matthias and Tkatchenko, Alexandre and Scheffler, Matthias},
  title   = {Hydrogen bonds and van der {Waals} forces in ice at ambient and high pressures},
  journal = {Physical Review Letters},
  year    = {2011},
  volume  = {107},
  number  = {18},
  pages   = {185701},
  doi     = {10.1103/PhysRevLett.107.185701}
}

@misc{batatia2023macehigherorderequivariant,
  author        = {Batatia, Ilyes and Kov{\'a}cs, D{\'a}vid P{\'e}ter and Simm, Gregor N. C. and Ortner, Christoph and Cs{\'a}nyi, G{\'a}bor},
  title         = {{MACE}: Higher order equivariant message passing neural networks for fast and accurate force fields},
  year          = {2023},
  eprint        = {2206.07697},
  archivePrefix = {arXiv},
  primaryClass  = {stat.ML}
}

@inproceedings{Batatia2022mace,
  author    = {Batatia, Ilyes and Kovacs, David Peter and Simm, Gregor N. C. and Ortner, Christoph and Csanyi, Gabor},
  title     = {{MACE}: Higher order equivariant message passing neural networks for fast and accurate force fields},
  booktitle = {Advances in Neural Information Processing Systems},
  editor    = {Oh, Alice H. and Agarwal, Alekh and Belgrave, Danielle and Cho, Kyunghyun},
  year      = {2022},
  url       = {https://openreview.net/forum?id=YPpSngE-ZU}
}

@article{batatia2025design,
  author  = {Batatia, Ilyes and Batzner, Simon and Kov{\'a}cs, D{\'a}vid P{\'e}ter and Musaelian, Albert and Simm, Gregor N. C. and Drautz, Ralf and Ortner, Christoph and Kozinsky, Boris and Cs{\'a}nyi, G{\'a}bor},
  title   = {The design space of {E(3)}-equivariant atom-centred interatomic potentials},
  journal = {Nature Machine Intelligence},
  year    = {2025},
  volume  = {7},
  number  = {1},
  pages   = {56--67},
  doi     = {10.1038/s42256-024-00956-x}
}

@misc{Li2025AbInitioMelting,
  author        = {Li, Yifan and Yang, Bingjia and Zhang, Chunyi and Gomez, Axel and Xie, Pinchen and Chen, Yixiao and Piaggi, Pablo M. and Car, Roberto},
  title         = {Ab initio melting properties of water and ice from machine learning potentials},
  year          = {2025},
  eprint        = {2512.23939},
  archivePrefix = {arXiv},
  primaryClass  = {cond-mat.mtrl-sci}
}

@article{ChewReinhardt2023,
  author  = {Chew, Pin Yu and Reinhardt, Aleks},
  title   = {Phase diagrams—why they matter and how to predict them},
  journal = {Journal of Chemical Physics},
  year    = {2023},
  volume  = {158},
  number  = {3},
  pages   = {030902},
  doi     = {10.1063/5.0131028}
}

\end{document}